\documentclass[showpacs,preprintnumbers,amsmath,amssymb]{revtex4}
\usepackage{amsmath,amssymb,graphics,epsfig,subfigure}
\usepackage{color}

\begin{document}
\renewcommand{\baselinestretch}{1.3}

\title{Effects of tidal charge on magnetic reconnection and energy extraction from spinning braneworld black hole}

\author{Shao-Wen Wei \footnote{weishw@lzu.edu.cn, corresponding author},
Hui-Min Wang \footnote{wanghm17@lzu.edu.cn},
Yu-Peng Zhang \footnote{zyp@lzu.edu.cn},
Yu-Xiao Liu \footnote{liuyx@lzu.edu.cn}}

\affiliation{Lanzhou Center for Theoretical Physics, Key Laboratory of Theoretical Physics of Gansu Province, School of Physical Science and Technology, Lanzhou University, Lanzhou 730000, People's Republic of China,\\
 Institute of Theoretical Physics $\&$ Research Center of Gravitation,
Lanzhou University, Lanzhou 730000, People's Republic of China.}

\begin{abstract}
Recent study shows that the magnetic reconnection can serve as an efficient mechanism to extract energy from rapidly spinning black holes. In this paper, we consider the effects of the tidal charge on the energy extraction via the magnetic reconnection in the backgrounds of a spinning braneworld black hole. With the increase of the tidal charge, we find that both the energies of the accelerated plasma and decelerated plasma decrease for the observer located at infinity. To achieve the purpose  extracting the net energy from the black hole, the decelerated plasma should take negative energy. Then we observe that the power of the energy extraction via the magnetic reconnection grows with the tidal charge. Meanwhile, the efficiency also increases with the tidal charge indicating that the spinning braneworld black hole with positive tidal charge is more efficient than the Kerr black hole. Compared with the Blandford-Znajek mechanism, the magnetic reconnection process shall have a higher power for positive tidal charge. These results indicate that the tidal charge has a significant effect on the energy extraction via the magnetic reconnection process.
\end{abstract}



\maketitle

\section{Introduction}

Black holes are the most intriguing astrophysical objects relating with highly energetic phenomena, such as black hole binary mergers \cite{Abbott}, active
galactic nuclei \cite{McKinney,Hawley,Komissarov,McKinneyb}, as well as the gamma-ray bursts \cite{Lee,Narayan,Barkov}. All them have an important role in the evolution of our universe. These enormous amounts of released energy can origin from the gravitational potential energy, electromagnetic field energy around the black hole, or the energy of the black hole itself. Understanding these energy extraction mechanisms has fundamental implications for these high energy astrophysical phenomena operated by the black holes.

For a spinning Kerr black hole characterized by the mass $M$ and spin $a$, Christodoulou \cite{Christodoulou} found that a portion of the black hole mass
\begin{eqnarray}
 M_{\text{irr}}=M\sqrt{\frac{1}{2}\left(1+\sqrt{1-\frac{a^2}{M^2}}\right)}
\end{eqnarray}
is irreducible, with which black hole horizon area is $A_{H}=16\pi M_{\text{irr}}^2$. According to the Beckenstein-Hawking entropy area law \cite{Bekensteina,Bekensteinb,Hawking}, the entropy $S=A/4=4\pi M_{\text{irr}}^2$. Thus the irreducible mass is consistent with the second law of black hole thermodynamics. On the other hand, without violating the second law, the maximum amount of energy that can be extracted is
\begin{eqnarray}
 E_{\text{rot}}=\left(1-\sqrt{\frac{1}{2}\left(1+\sqrt{1-\frac{a^2}{M^2}}\right)}\right)M.
\end{eqnarray}
For a non-spinning black hole $a/M$=0, one easily gets $E_{\text{rot}}$=0, so this portion energy should be the rotational energy of the black hole. It reaches the maximal value $E_{\text{rot}}\approx0.29M$ for an extremal spinning Kerr black hole with $a/M$=1.

Such irreducible mass presents an upper bound for the extractable rotational energy. However, the first realized mechanism was proposed by Penrose \cite{Penrose}, and now it is well-known as the Penrose process. The seminal thought experiment is the particle fission, that one falling particle splits into two (i.e. 0$\rightarrow$1+2). These two particles have different directions with the black hole spin. For example, particle 1 is absorbed by the black hole, while the particle 2 escapes to infinity. In general, the energy of particle 2 is smaller than the initial particle 0 according to the conservation of energy. However, if this particle fission occurs in the ergosphere of the black hole, the result shall be quite different. One particular property of the ergosphere is that this region allows the existence of a negative energy particle seen by an observer located at infinity, nevertheless a particle always has positive energy. Holding this idea, if the absorbed particle 1 has negative energy in the ergosphere, then the escaping particle 2 must have energy larger than the particle 0. At the same time, the black hole energy decreases after it absorbs particle 1. As a result, extra energy is extracted from the black hole. Energy extraction was firstly by this Penrose process and after that the interest in this field was sparked.

Although original Penrose process shed new light on the energy extraction from a spinning black hole, it is hard to implement in astrophysical scenarios. The main reason is that, during this process, these two newborn particles need to have a very high separation speed, which is approximately greater than half of the speed of light \cite{Bardeen,Wald}. The expected event rate is also very rare. So extracting a considerable amount of rotational energy is extremely hard. Subsequently, seeking the improved mechanisms to extract energy becomes interesting. Superradiant scattering \cite{Teukolsky}, the particles accelerators \cite{Banados,Wei}, collisional Penrose process \cite{Piran}, the
Blandford-Znajek (BZ) process \cite{Blandford}, and the magnetohydrodynamic Penrose process \cite{Takahashi} are the several well known alternative mechanisms. Especially, the BZ process is believed to be the leading mechanism powering the relativistic jets of active galactic nuclei \cite{McKinney,Hawley,Komissarov,McKinneyb} and the gamma-ray bursts \cite{Lee,Narayan,Barkov} via the environmental magnetic field around black hole.

In an astrophysical scenario, magnetic fields are thought to extensively exist near the black holes. Recent EHT collaboration's observation \cite{Akiyama,Akiyamab} of the polarized emission for the supermassive black hole M87* prominently confirmed this point. So considering energy extraction through magnetic fields is a promising approach. Towards to this study, the energy extraction through the rapid magnetic reconnection attracts attention. During the magnetic reconnection, the angular momentum of the plasma particles will be redistributed, and thus it has a probability to extract the energy from the black hole. In Ref. \cite{Koide}, Koide and Arai exploratively studied the feasibility conditions for energy extraction by using the outflow jets produced in a laminar reconnection configuration with a purely toroidal magnetic field. The results indicated that relativistic reconnection is necessary  for extracting energy. Compared with the BZ process, the general-relativistic kinetic simulations of black hole magnetospheres \cite{Parfrey} indicate that the accelerated plasma particles may spread onto negative energy-at-infinity trajectories and contribute  to black hole energy extraction in a variant of the Penrose process. Very recently, a big progress was made by Comisso and Asenjo \cite{Comisso}. They considered the energy extraction through the magnetic reconnection in the ergosphere of rapid spinning Kerr black holes. The power and the efficiency were evaluated. The results show that the rapidly spinning Kerr black holes are viable sources for the magnetic reconnection to extract black hole energy. And thus this provides us a new mechanism to extract black hole energy. Soon, such study was generalized to the spinning black hole with broken Lorentz symmetry \cite{Khodadi}. It reveals that the negative Lorentz symmetry breaking parameter increases the power and efficiency of the energy extraction via the fast magnetic reconnection. As a result, extracting energy from a spinning black hole in modified gravity may be more promising than from its counterparts in general relativity.

On the other hand, the extra dimension theory and braneworld scenario have attracted particular interest, not only in particle physics and string theory but also in astronomical observations. Recently, observations of the gravitational waves via black hole binary merger by LIGO and Virgo collaboration and black hole image from the EHT collaboration have been used to test and constrain extra dimensions \cite{Yu,Andriot,Bolis,Vagnozzi,Banerjee}. The four-dimensional effective gravitational field equations can be obtained by projecting the bulk Einstein tensor on the brane hypersurface. It contains the four-dimensional Einstein tensor without the extra spatial dimensions and the projection of the bulk Weyl tensor onto the brane hypersurface. Since the projected Weyl tensor is traceless, it mimics the energy-momentum tensor of a Maxwell field. The static and spherically symmetric vacuum brane spacetime solution was obtained in Ref.~\cite{Dadhich}. And the stationary and axisymmetric counterpart was given in Ref.~\cite{Gumrukcuoglu}. Different from the electric charge which appears in the form of square in the solution, the tidal charge here can either be positive or negative. Such tidal charge was constrained via the astronomical observations and served as an indicator of extra dimensions \cite{Vagnozzi,Banerjee}. In particular, energy extraction from the braneworld black holes has attracted much attention. The phenomenon of superradiance was considered in Ref. \cite{Oliveira}. By computing the amplification factors of massless scalar waves, it was found that the tidal charge considerably enhances the energy extraction from near-extreme black holes. Particle collisions near the black holes also provide another mechanism to extract black hole energy. Considering that two infalling particles collide near the event horizon of the braneworld black hole, the ultra-high centre-of-mass energy will occur if some fine tuning conditions are satisfied \cite{Blaschke}.  The collisional Penrose process in ergoregion of the braneworld Kerr black hole was also studied in Ref. \cite{Khan}. The efficiency of energy extraction from the brane Kerr black hole was obtained, and it increased with the tidal charge. For the massive spinning particles, this collisional Penrose process was also investigated \cite{DuDu}. Negative tidal charge will produce a higher efficiency than its Kerr black hole counterparts. All these results indicate that the tidal charge of the braneworld black hole has an important role in the extraction of black hole energy. As shown by Comisso and Asenjo \cite{Comisso}, magnetic reconnection acts as an efficient energy extraction mechanism. So in this paper, we would like to consider the magnetic reconnection and energy extraction in the ergoregion of the spinning braneworld black hole given in Randall-Sundrum braneworld scenario \cite{Gumrukcuoglu}, and to study the effects of the tidal charge on the energy extraction process.

The present paper is structured as follows. In Sec. \ref{iner}, we consider the geodesics of a particle around the spinning braneworld black hole. Several characteristic orbit radii will be obtained, which are closely related to the energy extraction. In Sec. \ref{eevrm}, we calculate the energy extracted via the magnetic reconnection process. The possible parameter regions are also explored. Effects of the tidal charge are carefully analyzed. The power and efficiency via magnetic reconnection are examined in Sec. \ref{paevmr}, where the result shows that the positive tidal charge will enhance them. Finally, we summarize and discuss our results in Sec. \ref{Conclusion}.

\section{Geodesics of a particle around the spinning braneworld black hole}
\label{iner}

In the Boyer-Lindquist coordinates, the rotating black hole in the Randall-Sundrum braneworld scenario has the following form \cite{Gumrukcuoglu}
\begin{eqnarray}
 ds^2=g_{tt}dt^2+2g_{t\phi}dtd\phi+g_{\phi\phi}d\phi^2+g_{rr}dr^2+g_{\theta\theta}d\theta^2,
\end{eqnarray}
where the nonvanishing metric components are
\begin{eqnarray}
 g_{tt}=\frac{2Mr-b}{\Sigma}-1,\quad
 g_{t\phi}=-\frac{a(2Mr-b)}{\Sigma}\sin^2\theta,\quad
 g_{\phi\phi}=\frac{A}{\Sigma}\sin^2\theta,\quad
 g_{rr}=\frac{\Sigma}{\Delta},\quad
 g_{\theta\theta}=\Sigma.
\end{eqnarray}
The metric functions read
\begin{eqnarray}
 \Delta=r^2-2Mr+a^2+b,\quad
 \Sigma=r^2+a^2\cos^2\theta,\quad
 A=(r^2+a^2)^2-a^2\Delta\sin^2\theta,
\end{eqnarray}
where $M$, $a$, and $b$ are black hole mass, spin and tidal charge, respectively. Solving $\Delta=0$, we can obtain the radii of the horizons, which are
\begin{eqnarray}
 r_{\pm}=M\pm\sqrt{M^2-a^2-b}.
\end{eqnarray}
When $M^2>a^2+b$, the system has two horizons and corresponds to a black hole. When $M^2=a^2+b$, there is only one degenerate horizon for the extremal black hole. While for $M^2<a^2+b$, it represents a naked singularity rather than a black hole. In this paper, we only deal with the black hole case requiring a maximal value of $b_{\text{max}}=M^2-a^2$. The ergospheres of the black hole are determined by $g_{tt}=0$. The outer one locates at
\begin{eqnarray}
 r_{\text{E}}=M+\sqrt{M^2-a^2\cos^2\theta-b}.
\end{eqnarray}
At the equatorial plane $\theta=\pi/2$, we always have $r_{\text{E}}>r_{+}$ for nonvanishing black hole spin. And the ergosphere region is $r_{\text{E}}>r>r_{+}$.

\begin{figure}
\center{\subfigure[]{\label{OrbitsR09a}
\includegraphics[width=7cm]{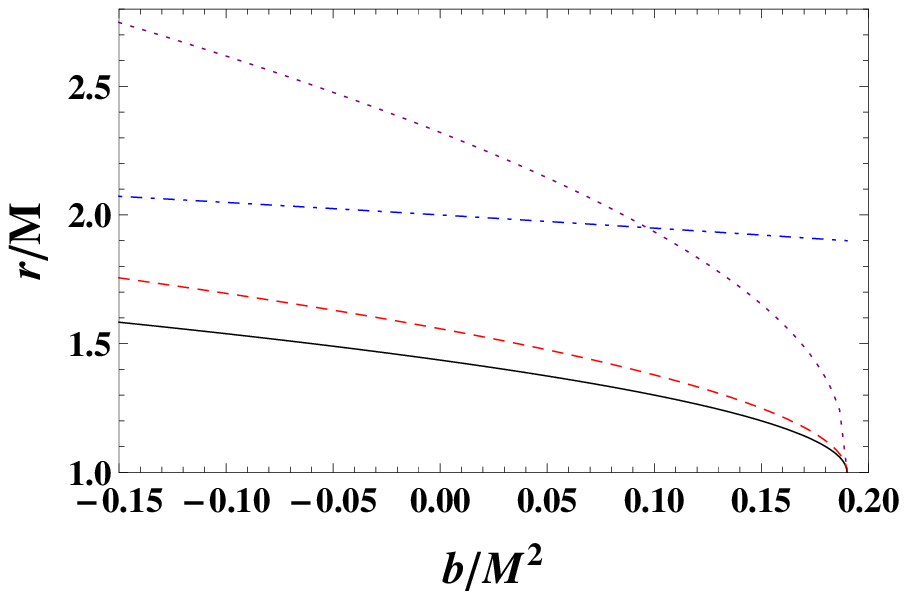}}
\subfigure[]{\label{OrbitsR099b}
\includegraphics[width=7cm]{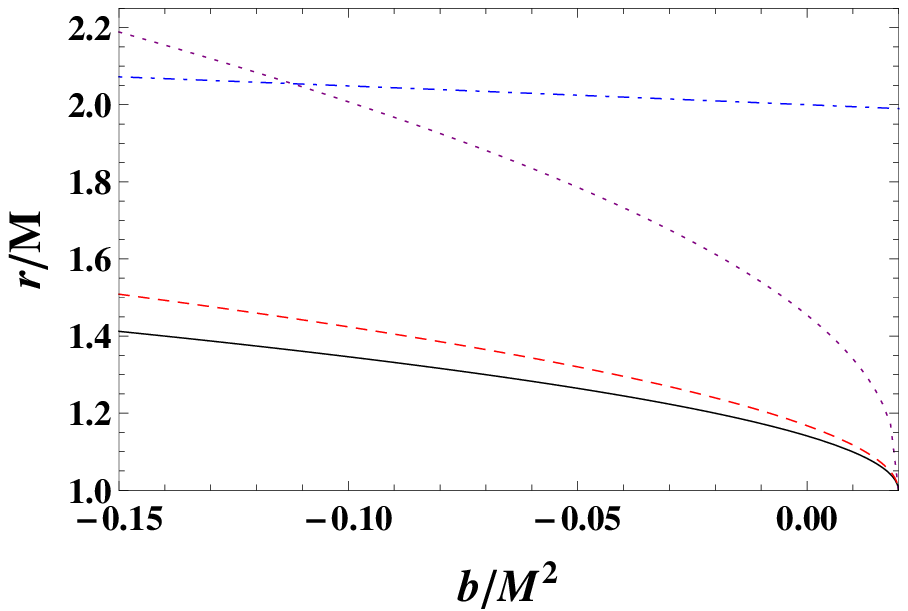}}}
\caption{Characteristic radii of the outer horizon (black solid curves), outer ergosphere (blue dot dashed curves), light ring (red dashed curves), and ISCO (purple dot curves) as functions of the tidal charge $b/M^2$. (a) $a/M$=0.9 with $b_{\text{max}}$=0.19. (b) $a/M$=0.99 with $b_{\text{max}}$=0.0199.}\label{ppOrbitsR099b}
\end{figure}

Using the Hamilton-Jacobi method, the geodesics of a particle around the rotating braneworld black hole takes the following forms \cite{Gumrukcuoglu}
\begin{eqnarray}
 \Sigma\frac{dr}{d\lambda}&=&\pm\sqrt{R},\label{radia}\quad\\
 \Sigma\frac{d\theta}{d\lambda}&=&\pm\sqrt{\Theta},\\
 \Sigma\frac{d\phi}{d\lambda}&=&-\frac{P_{\theta}}{\sin^2\theta}+\frac{aP_{r}}{\Delta},\\
 \Sigma\frac{dt}{d\lambda}&=&-aP_{\theta}+\frac{(r^2+a^2)P_{r}}{\Delta},
\end{eqnarray}
where
\begin{eqnarray}
 P_{\theta}&=&aE\sin^2\theta-L,\\
 P_r&=&E(r^2+a^2)-aL,\\
 R&=&P^2_{r}-\Delta(m^2r^2+Q+(aE-L)^2),\\
 \Theta&=&Q+(aE-L)^2-a^2m^2\cos^2\theta-\frac{P_{\theta}^2}{\sin^2\theta}.
\end{eqnarray}
The parameter $Q$ is the Cater constant and $\lambda$ is the affine parameter along the geodesics. And $m$, $E$, and $L$ are, respectively, the rest mass, energy, and angular momentum of the particle around the black hole. In particular for photon and massive particle, we can adopt $m^2$=0 and 1 for simplicity. In this paper we simplify the assumption that magnetic reconnection takes place in the bulk plasma that circularly rotates around the black hole at the equatorial plane, so we only limit our attention on these circular orbits with $\theta=\pi/2$. The Keplerian angular velocity denotes that of a particle orbiting along the circular orbit of radius $r$ is
\begin{eqnarray}
 \Omega_{\text{K}}=\frac{d\phi/d\lambda}{dt/d\lambda}
 =\frac{aE(b-2Mr)-L(b+r(r-2 M))}{a^2E (b-r (2 M+r))-aL(b-2Mr)-Er^4}.\label{omegav}
\end{eqnarray}
However, here $E$, $L$, and $r$ are not independent. In order to explore their relation, we can rewrite the radial equation (\ref{radia}) as
\begin{eqnarray}
 \left(\Sigma\frac{dr}{d\lambda}\right)^2+V_{\text{eff}}=0,
\end{eqnarray}
with $V_{\text{eff}}=-R$. A particle along the circular orbit should have vanishing radial velocity and acceleration, which give the conditions
\begin{eqnarray}
 R=0,\quad \partial_{r}R=0.
\end{eqnarray}
Solving them, we can obtain energy $E$ and angular momentum $L$ in term of $r$ \cite{Gumrukcuoglu}
\begin{eqnarray}
 E&=&\frac{r^2+b-2Mr\pm a\sqrt{Mr-b}}{r\sqrt{r^2+2b-3Mr\pm 2a\sqrt{Mr-b}}},\\
 L&=&\pm\frac{\sqrt{Mr-b}(r^2+a^2\mp2a\sqrt{Mr-b})\mp ba}{r\sqrt{r^2+2b-3Mr\pm 2a\sqrt{Mr-b}}}.
\end{eqnarray}
After plugging them into (\ref{omegav}), we have the Keplerian angular velocity
\begin{eqnarray}
 \Omega_{\text{K}}=\pm\frac{\sqrt{Mr-b}}{r^2\pm a\sqrt{Mr-b}}.\label{Keplav}
\end{eqnarray}
The upper and lower signs refer to corotating and counterrotating orbits. Here we only concern the corotating orbits for the magnetic reconnection occurring inside the ergosphere. Such stable orbit can exist from infinity down to the innermost stable circular orbit (ISCO) with radius $r_{\text{ISCO}}$, which can be explicitly obtained by solving
\begin{eqnarray}
 R(r_{\text{ISCO}})=0,\quad \partial_{r}R(r_{\text{ISCO}})=0, \quad \partial_{r,r}R(r_{\text{ISCO}})=0.
\end{eqnarray}
While the radius of the unstable circular orbit can take values lower than it. Another limiting orbit is the unstable circular photon orbit, or light ring in the equatorial plane, which requires
\begin{eqnarray}
 R(r_{\text{LR}})=0,\quad \partial_{r}R(r_{\text{LR}})=0,\quad \partial_{r,r}R(r_{\text{LR}})>0.
\end{eqnarray}
Note that $m^2=0$ for a photon. We plot the characteristic radii of the outer horizon (black solid curves), outer ergosphere (blue dot dashed curves), light ring (red dashed curves), and ISCO (purple dot curves) as functions of the tidal charge $b/M^2$ for $a/M$=0.9 and 0.99 in Fig. \ref{ppOrbitsR099b}. With the increase of $b/M^2$, it is easy to find that the radii of the outer horizon, light ring, and ISCO decrease. And when the maximal value of $b_{\text{max}}^2$ is reached, these three radii tend to one. The radius of the outer ergosphere slightly decreases with $b/M^2$. On the other hand, all these characteristic radii decrease with the black hole spin. In this paper, we consider that the magnetic reconnection happens in the ergosphere region between the black solid curves and the blue dot dashed curves. This region enlarges with $b/M^2$ for each fixed black hole spin.

\section{Energy extraction via magnetic reconnection mechanism}
\label{eevrm}

In the equatorial plane, the fast magnetic reconnection is caused by the
formation of flux ropes/plasmoids \cite{Albright,Huang} when the current
sheet exceeds a critical aspect ratio \cite{Lingama,Uzdensky,Lingamb}. The available magnetic energy will convert into plasma particle energy, and thus the particle can come out of the reconnection layer \cite{Huang}. A large number of numerical simulations \cite{Parfrey,Ripperda,Bransgrove} demonstrated that there always exists a dominant X-point, where the magnetic field lines reconnect, and the plasma including the flux ropes/plasmoids are expelled \cite{Vishniacd}. In this magnetic reconnection mechanism, the accelerated and decelerated parts of plasma particles move in opposite directions in the black hole ergosphere. The decelerated particles move toward to the black hole and then are absorbed. For an observer located at infinity, these particles have negative energy, and thus black hole energy decreases. While these accelerated particles will carry out more energy from the black hole. Such process will be repeated again and again by the frame-dragging effect for a rapidly spinning black hole. Moreover, the reconnection location, the dominant X-point, is also very important for the energy extraction. In order to meet the requirement that the decelerated particles have negative energy for distant observer, the X-point must locate in the regime between the outer horizon and ergosphere of the black hole.

Now we turn to analyze the plasma energy density. For convenience, we adopt the locally nonrotating frame, the zero-angular-momentum-observer (ZAMO) frame. In this frame, the line element is cast in a new form
\begin{eqnarray}
 ds^2=-d\hat{t}^2+\sum_{i=1}^3(d\hat{x}^{i})^2=\eta_{\mu\nu}dx^{\mu}dx^{\nu}.
\end{eqnarray}
The coordinate transformation between these two frames is
\begin{eqnarray}
 d\hat{t}=\alpha dt,\quad d\hat{x}^{i}=\sqrt{g_{ii}}dx^{i}-\alpha\beta^{i}dt.
\end{eqnarray}
The lapse function $\alpha$ and shift vector $\beta^{i}=(0, 0, \beta^{\phi})$ are given by
\begin{eqnarray}
 \alpha=\sqrt{-g_{tt}+\frac{g_{\phi t}^2}{g_{\phi\phi}}},\quad
 \beta^{\phi}=\frac{\sqrt{g_{\phi\phi}}\omega^{\phi}}{\alpha},
\end{eqnarray}
where the angular velocity of the frame dragging is $\omega^{\phi}=-g_{\phi t}/g_{\phi\phi}$. The contravariant and covariant components of the vector $\psi$ in the ZAMO frame and Boyer-Lindquist coordinates are related by
\begin{eqnarray}
 \hat{\psi}^0&=&\alpha \psi^0,\quad \hat{\psi}^{i}=\sqrt{g_{ii}}\psi^{i}-\alpha\beta^{i}\psi^0,\label{vvect}\\
 \hat{\psi}_0&=&\frac{\psi_0}{\alpha}+\sum_{i=1}^3\frac{\beta^i}{\sqrt{g_{ii}}}\psi_i,\quad
 \hat{\psi}_i=\frac{\psi_i}{\sqrt{g_{ii}}}.
\end{eqnarray}
Here, we turn to evaluate the ability of the magnetic reconnection happening in the ergosphere region to extract black hole energy. The requirements are the plasma with negative energy is decelerated by the reconnection and then absorbed by the black hole, while the plasma with positive energy is accelerated and escapes far away from the black hole seen by the observer at infinity. In the one-fluid approximation, the energy-momentum tensor for the plasma can be expressed as
\begin{eqnarray}
 T^{\mu\nu}=pg^{\mu\nu}+wU^{\mu}U^{\nu}+F^{\mu}_{\;\;\;\delta}F^{\nu\delta}
 -\frac{1}{4}g^{\mu\nu}F^{\rho\delta}F_{\rho\delta},
\end{eqnarray}
with $p$, $w$, $U^{\mu}$, and $F^{\mu\nu}$ being the proper plasma pressure,
enthalpy density, four-velocity, and electromagnetic field tensor. The ``energy-at-infinity" density is defined as $e^\infty=-\alpha g_{\mu 0}T^{\mu 0}$, which gives
\begin{eqnarray}
 e^\infty=\alpha \hat{e}+\alpha \beta^{\phi}\hat{P}^\phi.
\end{eqnarray}
The total energy density $\hat{e}$ and azimuthal component $\hat{P}^{\phi}$ of the momentum density are
\begin{eqnarray}
 \hat{e}&=&w\hat{\gamma}^2-p+\frac{\hat{B}^2+\hat{E}^2}{2},\\
 \hat{P}^{\phi}&=&w\hat{\gamma}^2\hat{v}^{\phi}+(\hat{B}\times\hat{E})^{\phi},
\end{eqnarray}
where the Lorentz factor $\hat{\gamma}=\hat{U}^0=\sqrt{1-\sum_{i=1}^{3}(d\hat{v}^i)^2}$ and the
components of magnetic and electric fields $\hat{B}^i=\epsilon^{ijk}\hat{F}_{jk}/2$ and $\hat{E}^i=\eta^{ij}\hat{F}_{j0}=\hat{F}_{i0}$. $\hat{v}^{\phi}$ is the azimuthal component of the out flow velocity of plasma for a ZAMO observer.

In general, the energy-at-infinity density $e^{\infty}$ can be separated
into two parts, the hydrodynamic component and electromagnetic component, i.e., $e^{\infty}=e^{\infty}_{\text{hyd}}+e^{\infty}_{\text{em}}$,
\begin{eqnarray}
 e^{\infty}_{\text{hyd}}&=&\alpha\hat{e}_{\text{hyd}}+\alpha\beta^{\phi}w\hat{\gamma}^2\hat{v}^{\phi},\\
 e^{\infty}_{\text{em}}&=&\alpha\hat{e}_{\text{em}}+\alpha\beta^{\phi}(\hat{B}\times\hat{E})_{\phi},
\end{eqnarray}
with $\hat{e}_{\text{hyd}}=w\hat{\gamma}^2-p$ and $\hat{e}_{\text{em}}=(\hat{B}^2+\hat{E}^2)/2$ denoting the hydrodynamic and electromagnetic energy densities observed in the ZAMO frame. The magnetic reconnection process converts the magnetic energy into kinetic energy of the plasma. If one expects it is very efficient, the electromagnetic energy will be negligible when comparing with the hydrodynamic energy-at-infinity. Then by making use of the approximation that the plasma is incompressible and adiabatic, the energy-at-infinity density can be well evaluated by \cite{Koide}
\begin{eqnarray}
 e^{\infty}=e^{\infty}_{\text{hyd}}=\alpha w\hat{\gamma}(1+\beta^{\phi}\hat{v}^{\phi})-\frac{\alpha p}{\hat{\gamma}}.
\end{eqnarray}
To assess the localized reconnection process, one can introduce the local rest frame $x'^{\mu}$=($x'^{0}$, $x'^{1}$, $x'^{2}$, $x'^{3}$) for the bulk plasma rotating the black hole with Keplerian angular velocity $\Omega_{\text{K}}$ in the equatorial plane. The direction $x'^{1}$ is chosen to be parallel to the radial direction $x^{1}=r$, and direction $x'^{3}$ to the azimuthal direction $x^{3}=\phi$. In the ZAMO frame, we can express the corotating Keplerian velocity as
\begin{eqnarray}
 \hat{v}_{\text{K}}&=&\frac{d\hat{x}^{\phi}}{d\hat{x}^{t}}=\frac{d\hat{x}^{\phi}/d\lambda}{d\hat{x}^{t}/d\lambda}
 =\frac{\sqrt{g_{\phi\phi}}dx^{\phi}/d\lambda-\alpha\beta^{\phi}dx^{t}/d\lambda}{\alpha dx^{t}/d\lambda}\nonumber\\
 &=&\frac{\sqrt{g_{\phi\phi}}}{\alpha}\Omega_{\text{K}}-\beta^{\phi},
\end{eqnarray}
where we have used (\ref{vvect}). After replacing $\Omega_{\text{K}}$ in the above equation, we shall obtain the explicit form of $\hat{v}_{\text{K}}$. The corresponding Lorentz factor $\hat{\gamma}_{\text{K}}=1/\sqrt{1-\hat{v}_{\text{K}}^2}$. Taking the ``relativistic adiabatic incomprehensible ball approach", the hydrodynamic energy-at-infinity per enthalpy of the plasma has the following form \cite{Comisso}
\begin{eqnarray}
 \epsilon_{\pm}^{\infty}=\alpha\hat{\gamma}_{\text{K}}\left((1+\beta^{\phi}\hat{v}_{\text{K}})\sqrt{1+\sigma_0}\pm\cos\xi(\beta^{\phi}+\hat{v}_{\text{K}})\sqrt{\sigma_0}
 -\frac{\sqrt{1+\sigma_0}\mp \cos\xi \hat{v}_{\text{K}}\sqrt{\sigma_0}}{4\hat{\gamma}^{2}_{\text{K}}(1+\sigma_0-\cos^2\xi\hat{v}^{2}_{\text{K}}\sigma)}\right),
\end{eqnarray}
where $\sigma_0=B_0^2/w_0$ is the plasma magnetization upstream of the reconnection layer, and $\xi$ is the orientation angle between the magnetic field lines and the azimuthal direction in the equatorial plane. Therefore, we can find that the energy-at-infinity per enthalpy of the plasma associated with the accelerated/decelerated ($\pm$) plasma is parametrized by five parameters ($a$, $b$, $r$, $\sigma_0$, $\xi$). The first two describe the nature of the black hole, and the rest ones are related to the magnetic reconnection and the matter disc near the black hole.

If one expects to extract the black hole energy via the magnetic reconnection, the decelerated plasma should have negative energy as measured at infinity, while the accelerated plasma has positive energy-at-infinity larger than its rest mass and thermal energy, which gives
\begin{eqnarray}
 \epsilon_{-}^{\infty}<0,\quad
 \Delta\epsilon_{+}^{\infty}
 = \epsilon_{+}^{\infty}
  -\left(1-\frac{\Gamma}{\Gamma-1}\frac{p}{w}\right)
  =\epsilon_{+}^{\infty}>0,\label{condds}
\end{eqnarray}
for a relativistically hot plasma with polytropic index $\Gamma$=4/3.

It is instructive to consider the extremal black hole limit, where $a/M\rightarrow\sqrt{1-b/M^2}$, $r/M\rightarrow 1$, and $\xi\rightarrow 0$. Since the energy extraction always corresponds to a high spin, we here consider $a/M^2>1/2$, or equivalently $b/M^2<3/4$. For this case, we obtain the following forms in the extremal black hole limit
\begin{eqnarray}
 \epsilon_{-}^{\infty}&=&\frac{\sqrt{(1-b/M^2)(\sigma_0+1)}-2 (1-b/M^2)
   \sqrt{\sigma_0}}{\sqrt{3-4 b/M^2}},\\
 \epsilon_{+}^{\infty}&=&\frac{\sqrt{(1-b/M^2)
   \left(\sigma_0+1\right)}+2 (1-b/M^2) \sqrt{\sigma_0}}{\sqrt{3-4 b/M^2}}.
\end{eqnarray}
Requiring the condition (\ref{condds}), we have
\begin{eqnarray}
 \sigma_0>\sigma_{\text{bound}}=\frac{1}{3-4b/M^2}.
\end{eqnarray}
For $b/M^2$=0, it is the result of the spinning Kerr black hole \cite{Comisso}. We also can find that positive tidal charge increases while negative tidal charge decreases with the bound $\sigma_{\text{bound}}$ by comparing with the spinning Kerr black hole counterpart. This also implies that extracting rotational energy via magnetic reconnection is possible for $\sigma_0$ below unity when $b/M^2<1/2$. On the other hand, if one expects to extract sizable amounts of energy, higher $\sigma_0$ values are necessary. Taking $\sigma_0\gg1$, we obtain
\begin{eqnarray}
 \epsilon_{-}^{\infty}&\sim&\frac{\left(\sqrt{1-b/M^2}-2(1-b/M^2)\right) \sqrt{\sigma
   _0}}{\sqrt{3-4 b/M^2}},\\
 \epsilon_{+}^{\infty}&\sim&\frac{\left(\sqrt{1-b/M^2}+2(1-b/M^2)\right) \sqrt{\sigma
   _0}}{\sqrt{3-4 b/M^2}}.
\end{eqnarray}
These relations provide us with the energy-at-infinity per enthalpy of the accelerated and decelerated plasmas in the maximal energy extraction parameter region.

\begin{figure}
\center{\subfigure[]{\label{epsilonb01a.eps}
\includegraphics[width=7cm]{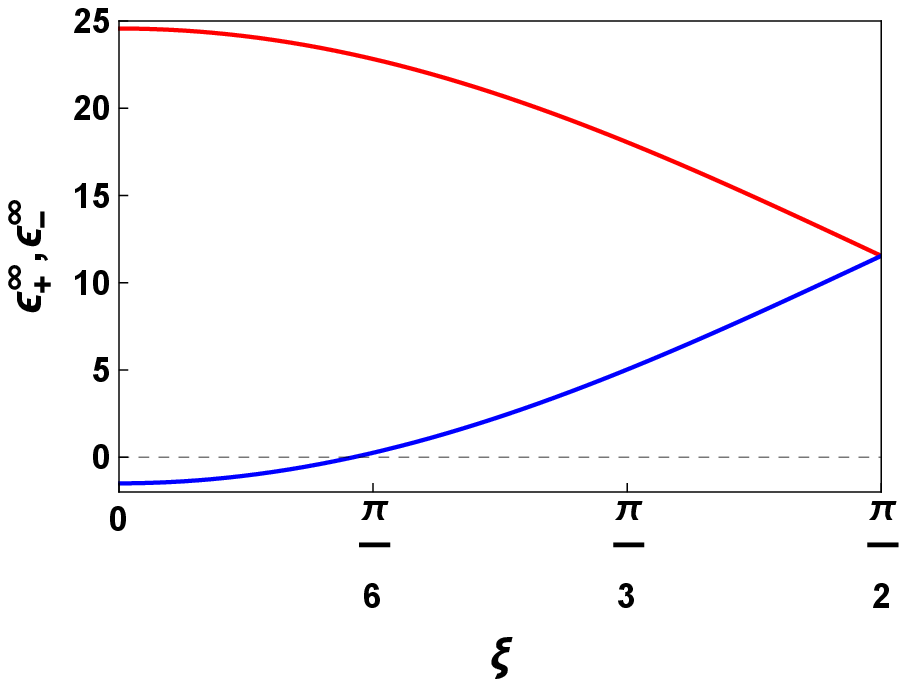}}
\subfigure[]{\label{epsilonb00b}
\includegraphics[width=7cm]{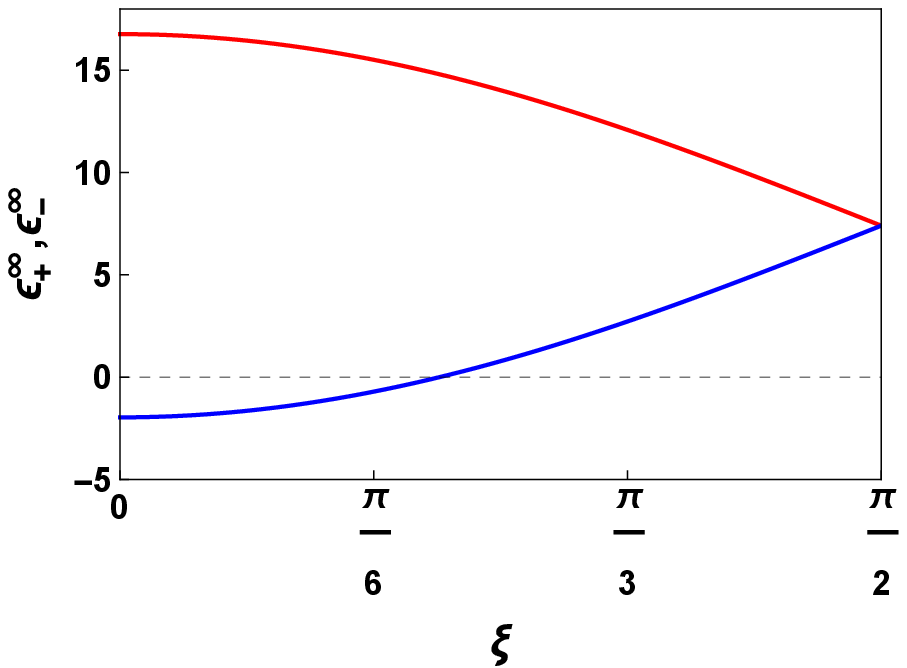}}}
\caption{The behaviors of $\epsilon_{+}^{\infty}$ (top curves) and $\epsilon_{-}^{\infty}$ (bottom curves) in terms of the orientation angle $\xi$ with $r/M$=1.5, $\sigma_0$=100, and $a/M$=0.99. (a) $b/M^2$=-0.1. (b) $b/M^2$=0.}\label{ppepsilonb00b}
\end{figure}

Now, we consider the effects of the orientation angle $\xi$ on the energy-at-infinity per enthalpy $\epsilon_{+}^{\infty}$ (solid curves) and $\epsilon_{-}^{\infty}$. Taking $r/M$=1.5, $\sigma_0$=100, and $a/M$=0.99, we plot them in Fig. \ref{ppepsilonb00b} for $b/M^2$=-0.1 and $b/M^2$=0. From the figures, it is easy to find that $\epsilon_{+}^{\infty}$ decreases, while $\epsilon_{-}^{\infty}$ increases with $\xi$. In order to extract the black hole energy, we require $\epsilon_{+}^{\infty}>0$ and $\epsilon_{-}^{\infty}<0$. The first one is satisfied for an arbitrary value of $\xi$. However, $\epsilon_{-}^{\infty}$ is negative for small $\xi$ and positive for large $\xi$. This suggests that low orientation angle $\xi$ is desirable for extracting energy through the magnetic reconnection process, and thus we will only discuss low orientation angle cases in the followings.

\begin{figure}
\center{\subfigure[]{\label{ETMBa90pi12a}
\includegraphics[width=7cm]{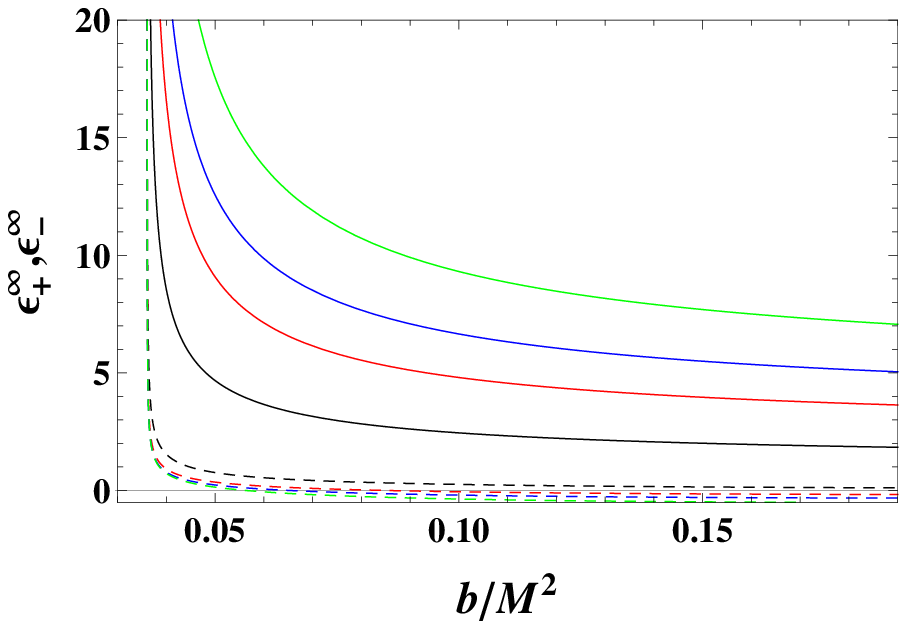}}
\subfigure[]{\label{ETMBa90pi6b}
\includegraphics[width=7cm]{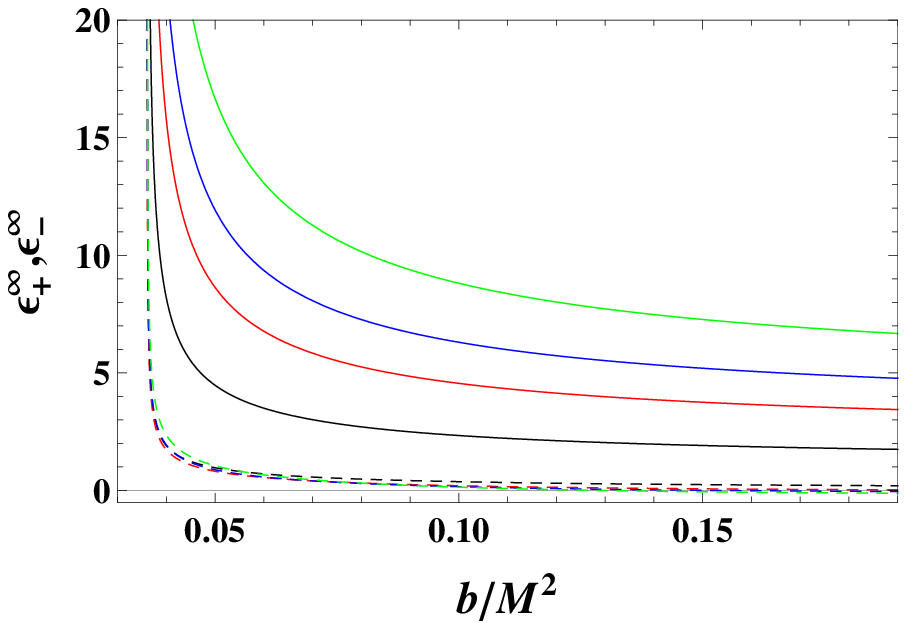}}
\subfigure[]{\label{ETMBa99pi12c}
\includegraphics[width=7cm]{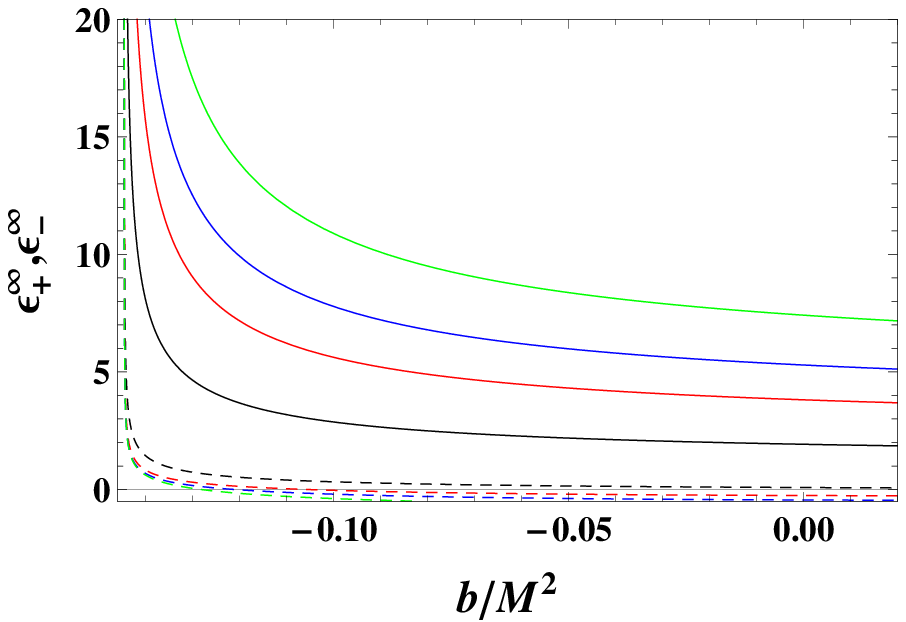}}
\subfigure[]{\label{ETMBa99pi6d}
\includegraphics[width=7cm]{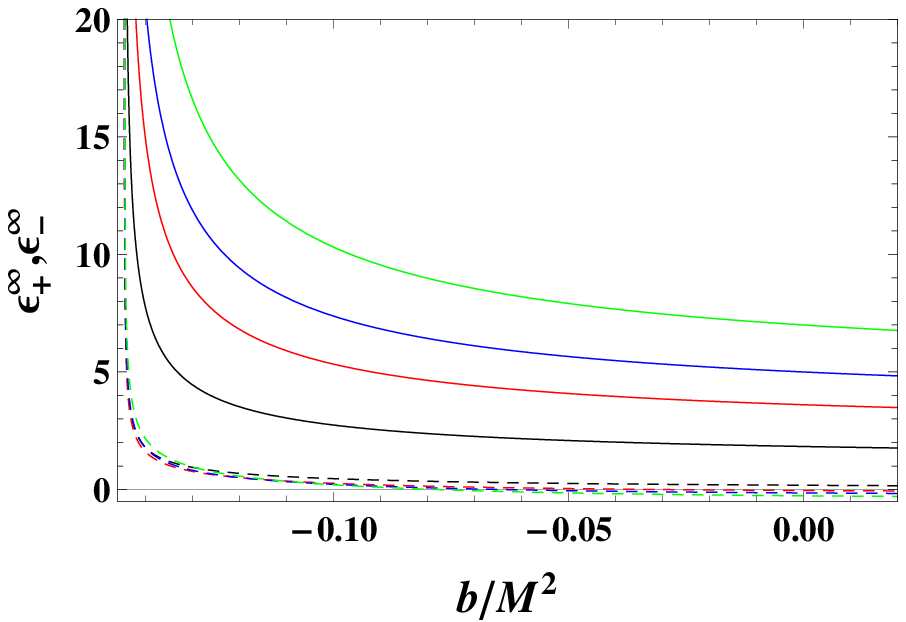}}}
\caption{The behaviors of $\epsilon_{+}^{\infty}$ (solid curves) and $\epsilon_{-}^{\infty}$ (dashed curves) in terms of the tidal charge $b/M^2$ with $r/M$=1.5. The plasma magnetization $\sigma_0$=1, 5, 10, and 20 from bottom to top for solid curves and from top to bottom for dashed curves. (a) $a/M$=0.9 and $\xi=\pi/12$. (b) $a/M$=0.9 and $\xi=\pi/6$. (c) $a/M$=0.99 and $\xi=\pi/12$. (d) $a/M$=0.99 and $\xi=\pi/6$.}\label{ppETMBa99pi6d}
\end{figure}

In order to examine the effects of the tidal charge on the energy-at-infinity per enthalpy, we plot $\epsilon_{+}^{\infty}$ (solid curves) and $\epsilon_{-}^{\infty}$ (dashed curves) as functions of the tidal charge $b/M^2$ for a given set of parameters with low $\xi$ in Fig.  \ref{ppETMBa99pi6d}. Although the spinning braneworld black hole can possess an arbitrarily negative tidal charge, both $\epsilon_{+}^{\infty}$ and $\epsilon_{-}^{\infty}$ start a finite negative value of the tidal charge $b/M^2$, and then decrease until $b_{\text{max}}$ is reached, where an extremal black hole is present. A detailed calculation shows that these starting points correspond to the tidal charge where the X-point meets the radius of the light ring. Comparing with $a/M$=0.9 and 0.99, it is easy to find that the starting point is shifted toward negative $b/M^2$ with the black hole spin. We can also find that $\epsilon_{+}^{\infty}$ is always positive while $\epsilon_{-}^{\infty}$ turns to negative at some certain values of the tidal charge $b/M^2$. Then it suggests the conditions (\ref{condds}) are satisfied and the energy extraction is available. For the same black hole spin, both $\epsilon_{+}^{\infty}$ and $\epsilon_{-}^{\infty}$ slightly increase with the orientation angle $\xi$. On the other hand, when other parameters are fixed, we observe that $\epsilon_{+}^{\infty}$ gains a significant increase when the magnetization parameter $\sigma_0$ takes values from 1 to 20, while $\epsilon_{-}^{\infty}$ slightly decreases.

\begin{figure}
\center{\subfigure[]{\label{ParaRgepi12a}
\includegraphics[width=7cm]{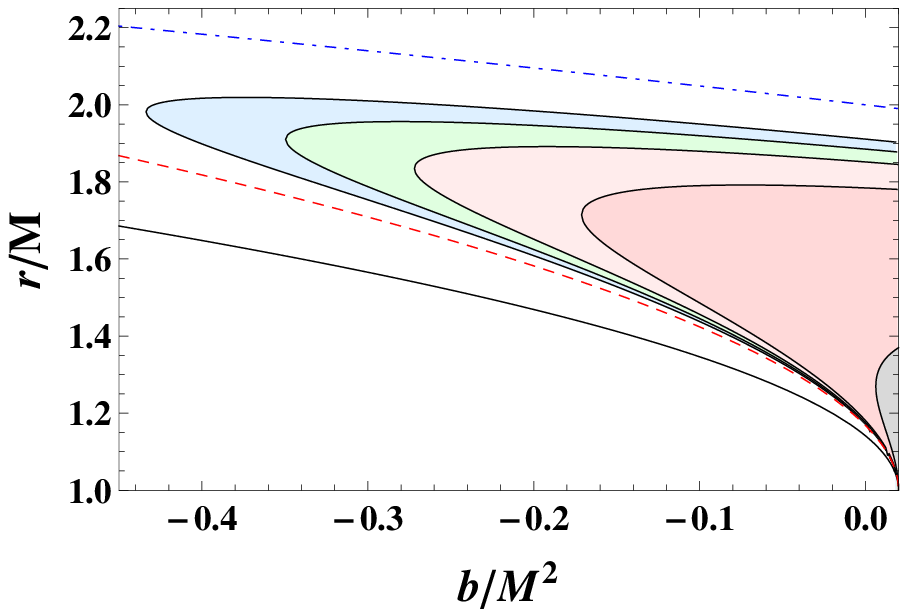}}
\subfigure[]{\label{ParaRgepi6}
\includegraphics[width=7cm]{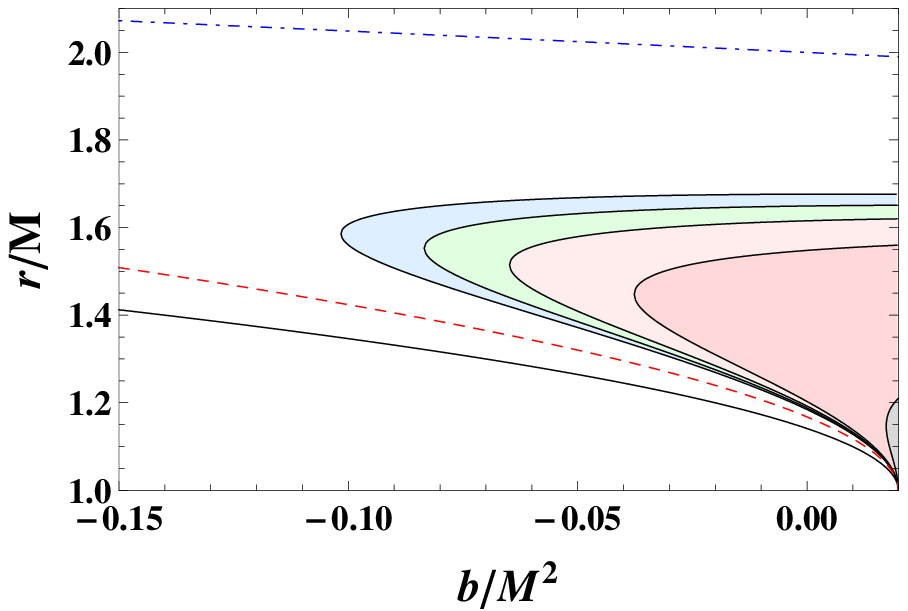}}}
\caption{Regions of the phase-space ($r/M$, $b/M^2$) with $a/M$=0.99. The color regions are for $\epsilon_{-}^{\infty}<0$ with $\sigma_0$=1, 5, 10, 20, 100 from right to left. Black solid curves, red dashed curves, and blue dot dashed curves denote the radii of the outer horizon, light ring, and outer ergosphere, respectively. (a) $\xi=\pi/12$. (b) $\xi=\pi/6$.}\label{ppParaRgepi6}
\end{figure}

Now, let us turn to analyze the parameter space for the conditions (\ref{condds}) with which the energy extraction via magnetic reconnection can be realized. In Ref. \cite{Comisso}, the authors suggested that the magnetic reconnection is a viable mechanism for energy extraction for a rapidly spinning black hole. Thus we here consider the black hole case with spin $a/M$=0.99. The results are displayed in the ($r/M$-$b/M^2$) parameter space in Fig. \ref{ppParaRgepi6} for the orientation angle $\xi=\pi/12$ and $\pi/6$ while the magnetization parameter $\sigma_0$=1, 5, 10, 20, and 100. In the unshaded parameter space, black solid curves, red dashed curves, and blue dot dashed curves are, respectively, for the radii $r_+$ of the outer black hole horizon, $r_{\text{LR}}$ of the light ring, and $r_E$ of the outer ergosphere.

As discussed above, $\Delta\epsilon_{+}^{\infty}$ requires the X-point locates in ($r_{\text{LR}}$, $r_{\text{E}}$), which is independent of $\xi$ and $\sigma_0$. While $\epsilon_{-}^{\infty}$ shall give a stronger constraint as expected. Here we mark the region of negative $\epsilon_{-}^{\infty}$ with different colors in the parameter space for different values of $\sigma_0$. Both for $\xi$=$\pi/12$ and $\pi/6$, the shaded regions for $\sigma_0$=1 are quite small, much near $b_{\text{max}}$. With the increase of $\sigma_0$, such regions grow toward to lower value of $b/M^2$ and larger value of $r/M$. When $\sigma_0$=100, we can see that the shaded regions of $\epsilon_{-}^{\infty}$ get a considerable expansion. Further increasing $\sigma_0$, the shaded region will extend up to the outer boundary of the ergosphere. Furthermore, comparing with Fig. \ref{ParaRgepi12a} and Fig. \ref{ParaRgepi6}, we easily observe that the region of $\epsilon_{-}^{\infty}$ shrinks with the orientation angle $\xi$.

In summary, for a fixed rapidly spinning black hole, we find that the parameter region that the energy can be extracted via magnetic reconnection enlarges with $\sigma_0$ while shrinks with $\xi$. Moreover, lower tidal charge requires large $\sigma_0$ and small $\xi$ to realize the energy extraction.

\section{Power and efficiency via magnetic reconnection}
\label{paevmr}

In this section, we aim to evaluate the power and efficiency of the energy extraction via the magnetic reconnection for a rapidly spinning braneworld black hole. The power and efficiency mainly depend on the amount of plasma of negative energy-at-infinity absorbed by the black hole in unit time. The power $P_{\text{e}}$ of the energy extraction from the black hole through the escaping plasma can be well estimated as \cite{Comisso}
\begin{eqnarray}
 P_{\text{e}}=-\epsilon_{-}^{\infty}w_0A_{\text{in}}U_{\text{in}},
\end{eqnarray}
where $U_{\text{in}}$=$\mathcal{O}(10^{-1})$ and $\mathcal{O}(10^{-2})$ for the collisionless and collisional regimes \cite{Comissojp}, respectively. For a rapidly spinning black hole, the cross sectional area of the inflowing plasma can be estimated as $A_{\text{in}}\sim(r_{\text{E}}^2-r_{\text{LR}}^2)$. For an extremal spinning braneworld black hole, one has $A_{\text{in}}\sim\sqrt{M^2-b}(2M+\sqrt{M^2-b})$.

\begin{figure}
\center{\subfigure[]{\label{Powersig100ra}
\includegraphics[width=7cm]{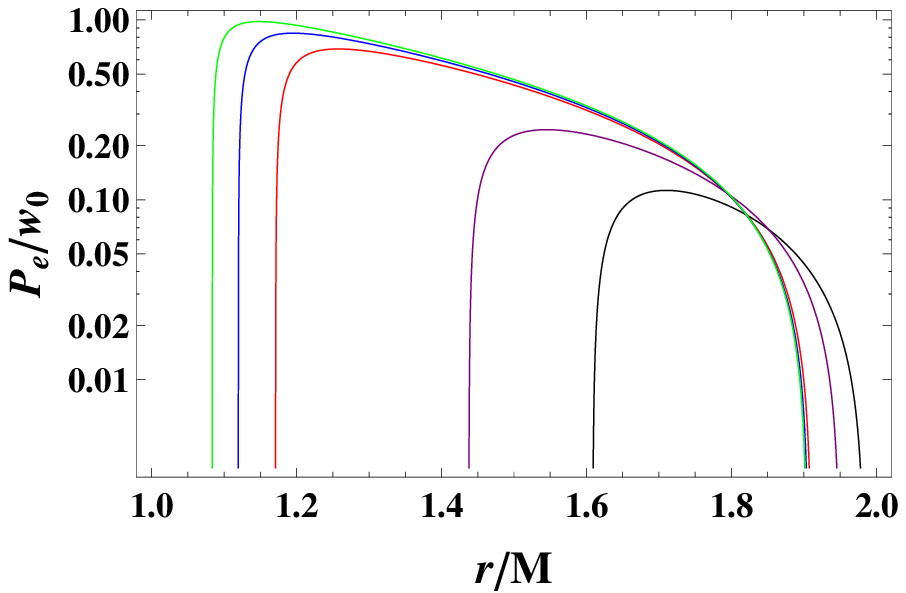}}
\subfigure[]{\label{Powersig10000rb}
\includegraphics[width=7cm]{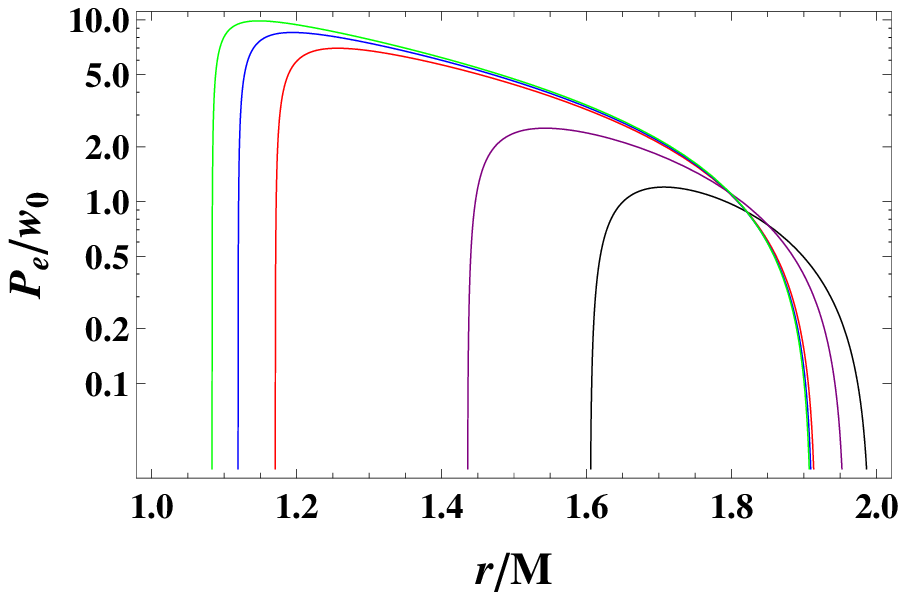}}}
\caption{Log-plot of the power $P_{\text{e}}/w_0$ as a function of the dominant X-point location $r/M$ for a rapidly spinning black hole with $a/M$=0.99. The reconnection inflow four-velocity $U_{\text{in}}$=0.1 for collisionless reconnection regime. The orientation angle $\xi$ is set to $\pi/12$. The tidal charge $b/M$=0.015, 0.01, 0, -0.1, -0.2 from left to right. (a) $\sigma_0$=100. (b) $\sigma_0$=10000. Note that the left points of these curves approach to the radii of the light rings.} \label{ppPowers}
\end{figure}

In Fig. \ref{ppPowers}, we plot the power $P_{\text{e}}/w_0$ as a function of the dominant X-point location $r/M$ for a rapidly spinning black hole with $a/M$=0.99. The reconnection inflow four-velocity $U_{\text{in}}$=0.1 and the orientation angle $\xi=\frac{\pi}{12}$. The tidal charge $b/M^2$=0.015, 0.01, 0, -0.1, -0.2 from left to right. And the magnetization parameter $\sigma_0$=100 in Fig. \ref{Powersig100ra} and 10000 in Fig. \ref{Powersig10000rb}. For different values of parameters, the power $P_{\text{e}}/w_0$ presents a similar pattern that a peak emerges at a certain X-point location. By decreasing the tidal charge $b/M^{2}$, the peak grows and is shifted at small $r/M$. In particular, after comparing with Fig. \ref{Powersig100ra} and Fig. \ref{Powersig10000rb}, we find that the peak gets a significant increase when $\sigma_0$ grows from 100 to 10000.

\begin{figure}
\center{\subfigure[]{\label{Powerrbm100a}
\includegraphics[width=7cm]{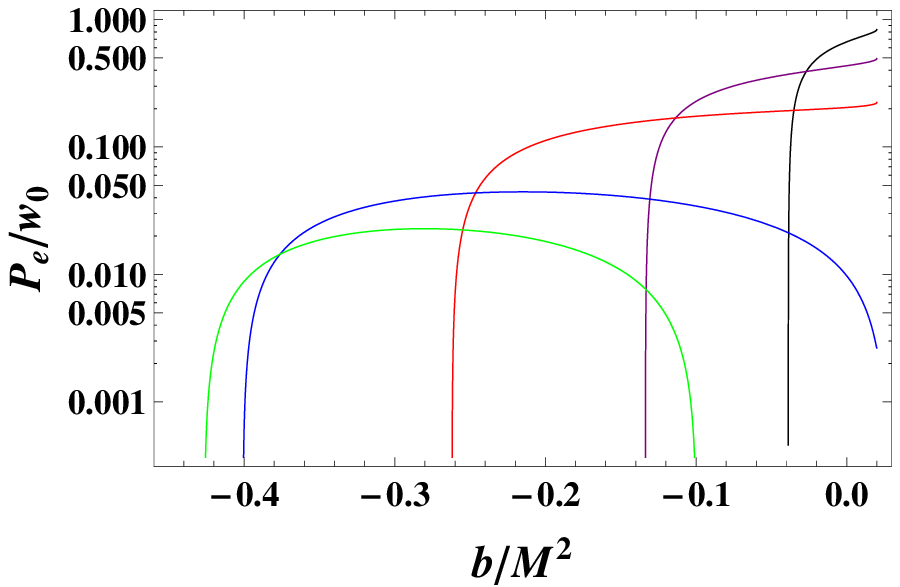}}
\subfigure[]{\label{Powerrbm10000b}
\includegraphics[width=7cm]{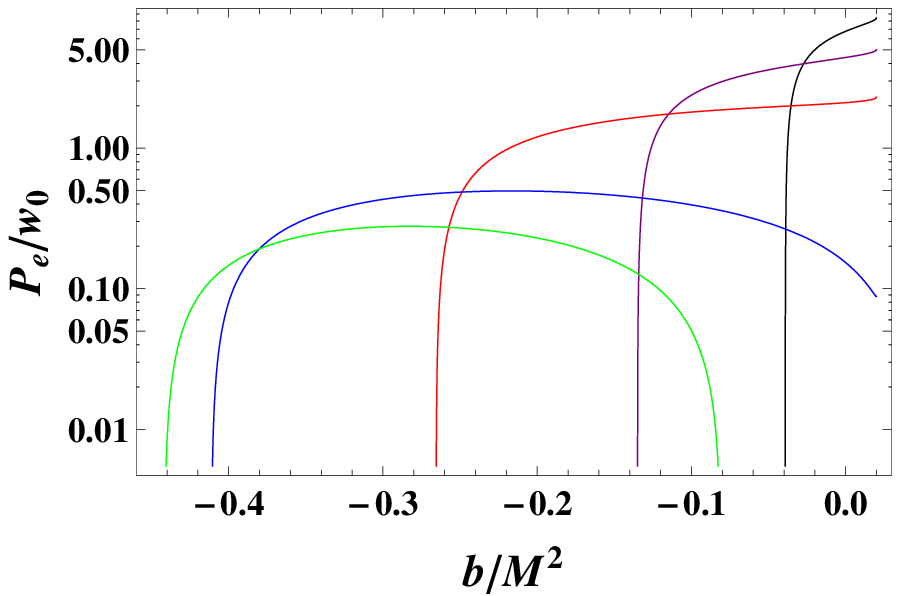}}}
\caption{Log-plot of the power $P_{\text{e}}/w_0$ as a function of the tidal charge $b/M^2$ for a rapidly spinning black hole with $a/M$=0.99. The reconnection inflow four-velocity $U_{\text{in}}$=0.1 for collisionless reconnection regime. The orientation angle $\xi$ is set to $\pi/12$. The dominant X-point location $r/M$=1.95, 1.9, 1.7, 1.5, 1.3 from left to right. (a) $\sigma_0$=100. (b) $\sigma_0$=10000.} \label{Powerrbm10000b}
\end{figure}

In order to show the effects of the tidal charge on the power, we show it in Fig. \ref{Powerrbm10000b} for $\sigma_0$=100 and 10000, respectively. The dominant X-point location $r/M$=1.95, 1.9, 1.7, 1.5, 1.3 from left to right. For both $\sigma_0$=100 and 10000, the power $P_{\text{e}}/w_0$ increases and then decreases with $b/M^2$ for larger X-point location $r/M$. When the location approaches to the black hole horizon, the pattern gradually changes. For example, when $r/M<1.7$, $P_{\text{e}}/w_0$ becomes a monotone increasing function with the tidal charge. The maximal power occurs at $b_{\text{max}}$=0.0199. Meanwhile, we observe that a higher $\sigma_0$ leads to a higher power.

By generating energetic plasma out flows via magnetic reconnection, energy extraction from a spinning black hole becomes possible. However, the key is the magnetic field energy, which can redistribute the angular momentum
of the particles such that the generated particles with negative energy-at-infinity are absorbed by the black hole and other positive energy particles escape
to infinity. Following Ref. \cite{Comisso}, we define the efficiency of the plasma energization process via magnetic reconnection as
\begin{eqnarray}
 \eta=\frac{\epsilon_{+}^{\infty}}{\epsilon_{+}^{\infty}+\epsilon_{-}^{\infty}}.
\end{eqnarray}
For the case of $\eta>1$, the energy will be extracted from the spinning black hole.

\begin{figure}
\center{\subfigure[]{\label{Effecrma}
\includegraphics[width=7cm]{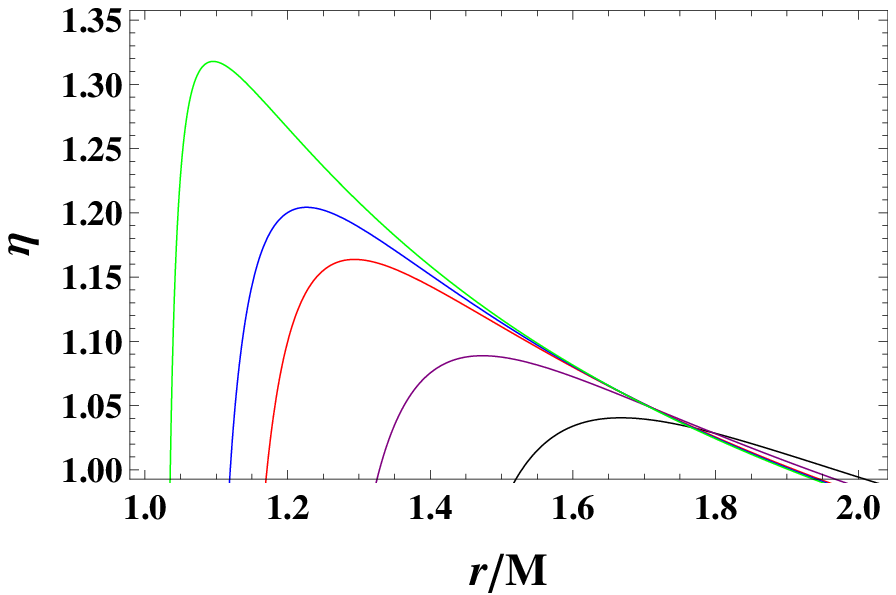}}
\subfigure[]{\label{Effecbb}
\includegraphics[width=7cm]{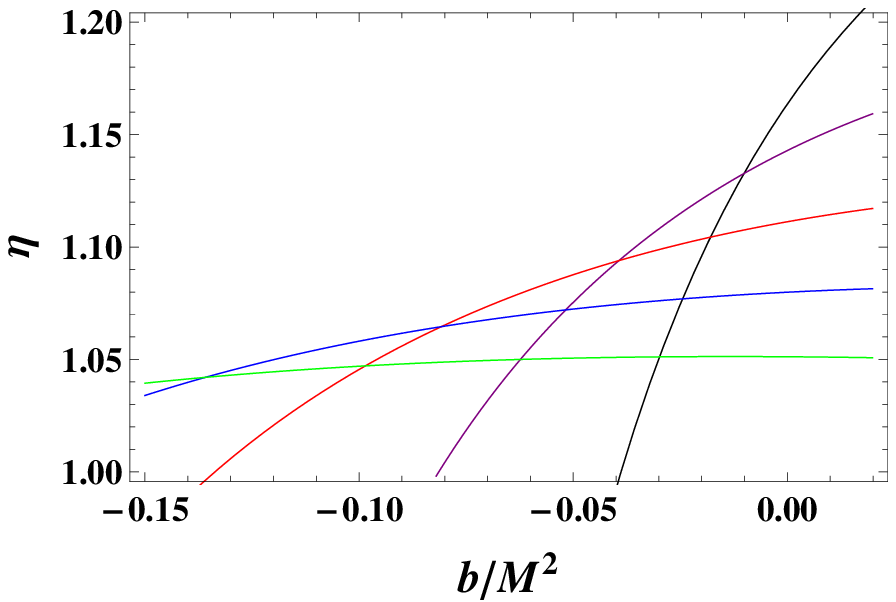}}}
\caption{Efficiency $\eta$ of the reconnection process for a rapidly spinning black hole with $a/M$=0.99. The parameters $\xi$ and $\sigma_0$ are set as $\xi=\frac{\pi}{12}$ and $\sigma_0$=100. (a) $\eta$ vs. $r/M$ for $b/M^2$=-0.15, -0.05, 0, 0.01, 0.019 from bottom to top. (b) $\eta$ vs. $b/M^2$ for $r/M$=1.3, 1.4, 1.5, 1.6, 1.7 from top to bottom in the right.} \label{PEffecbb}
\end{figure}

We show the efficiency $\eta$ in Fig. \ref{PEffecbb} for a rapidly spinning black hole with $a/M$=0.99, $\xi=\frac{\pi}{12}$, and $\sigma_0$=100. In Fig. \ref{Effecrma}, the efficiency $\eta$ is plotted as a function of X-point location $r/M$. The tidal charge is set to $b/M^2$=-0.15, -0.05, 0, 0.01, 0.019 from bottom to top. Obviously, the efficiency takes the maximal values at a moderate $r/M$. With the increase of $b/M^2$, the peak of the efficiency grows and is shifted to small $r/M$. In order to clearly show the effects of the tidal charge, we plot the efficiency in Fig. \ref{Effecbb}. With different values of $r/M$, the efficiency monotonically increases with the tidal charge. For large $r/M$, the efficiency slightly increases with the tidal charge $b/M^2$. While for small $r/M$, the efficiency gets a significant increase. Overall, we observe that the tidal charge will increase the efficiency of energy extraction via the magnetic reconnection.

Now we would like to make a comparison of the energy extraction via the magnetic reconnection and the BZ mechanism. From the BZ mechanism, the rate of energy extraction by including the higher order terms of the angular velocity $\Omega_{\text{H}}$ of the event horizon is given by \cite{Tchekhovskoyaj}
\begin{eqnarray}
 P_{\text{BZ}}=\frac{\kappa}{16\pi} \Phi^2_{\text{BH}} \Omega_{\text{H}}^2
    \left(1+\delta_1 \Omega_{\text{H}}^2+\delta_2\Omega_{\text{H}}^4+\mathcal{O}(\Omega_{\text{H}}^6)
    \right),
\end{eqnarray}
where $\delta_1$ and $\delta_2$ are the numerical coefficients and $\kappa$ is a numerical constant related to the magnetic field configuration \cite{Pei}. The angular velocity $\Omega_{\text{H}}=a/(r_{+}^{2}+a^2)$ for the spinning braneworld black hole. The magnetic flux threading one hemisphere of the black hole horizon is $\Phi_{\text{BH}}=\frac{1}{2}\int_{\theta}\int_{\phi}|B^r|dA_{\theta\phi}\sim B^{r}A_{H}/2\sim 2\pi(r_{+}^{2}+a^2)B_0\sin\xi$. Then it is natural to define the ratio between the power of these two mechanisms
\begin{eqnarray}
 \frac{P_{\text{e}}}{P_{\text{BZ}}}=
  \frac{-4\epsilon_{-}^{\infty}A_{\text{in}}U_{\text{in}}}{\pi\kappa\sigma_0(a^2+r_{+}^2)^2 \Omega_{\text{H}}^2 \sin^2\xi(1+\delta_1 \Omega_{\text{H}}^2+\delta_2\Omega_{\text{H}}^4)}.
\end{eqnarray}

\begin{figure}
\center{\subfigure[]{\label{Ratiosigmaa}
\includegraphics[width=7cm]{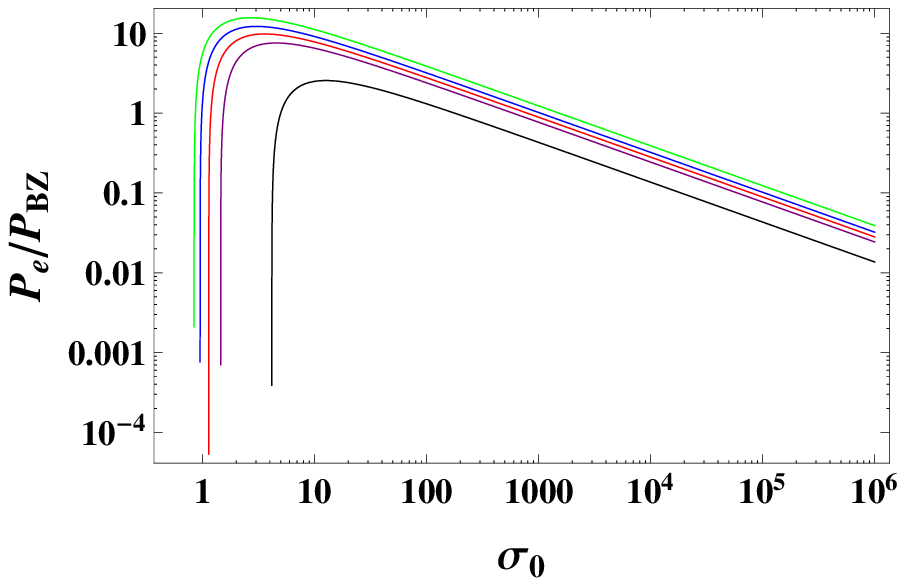}}
\subfigure[]{\label{Ratiobmb}
\includegraphics[width=7cm]{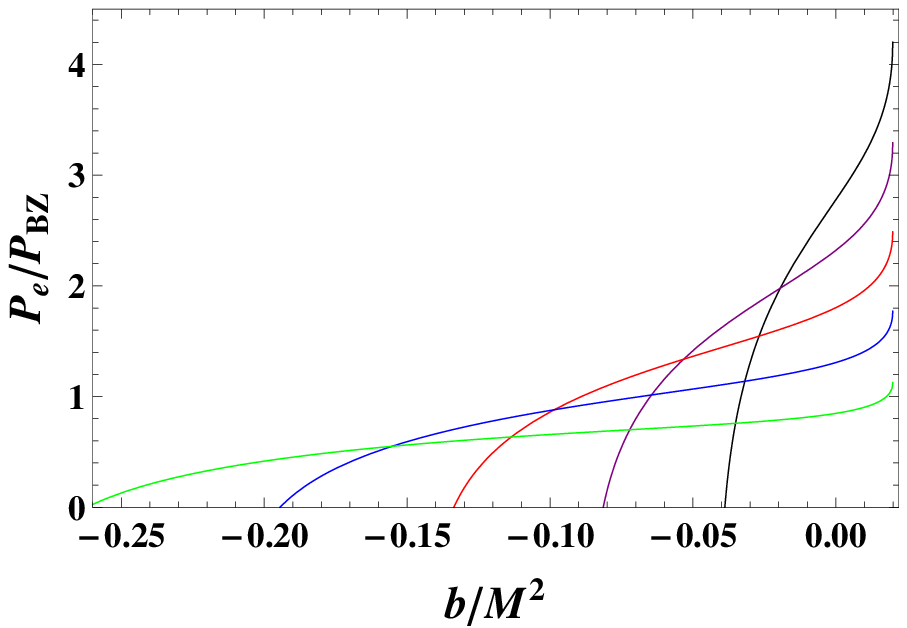}}}
\caption{Power ratio $P_{\text{e}}/P_{\text{BZ}}$ for a rapidly spinning black hole with $a/M$=0.99 and $\xi=\frac{\pi}{12}$. The parameters are set as $U_{\text{in}}$=0.1, $\kappa$=0.05, $\delta_1$=1.38, and $\delta_2$=-9.2 \cite{Tchekhovskoyaj}. (a) Log-Log plot of $P_{\text{e}}/P_{\text{BZ}}$ as a function of $\sigma_0$ for $r/M$=1.3 and $b/M^2$=-0.03, -0.01, 0, 0.01, 0.019 from bottom to top. (b) $P_{\text{e}}/P_{\text{BZ}}$ as a function of $b/M^2$ for $\sigma_0$=100 and $r/M$=1.3, 1.4, 1.5, 1.6, 1.7 from right to left.} \label{PRatiobmb}
\end{figure}

The ratio of the power is shown in Fig. \ref{PRatiobmb} for a rapidly spinning black hole with $a/M$=0.99 and $\xi=\frac{\pi}{12}$. In Fig. \ref{Ratiosigmaa}, we plot the ratio as a function of $\sigma_0$. The ratio rapidly increases with $\sigma_0$. After its maximal value is reached, it then approximately linearly decreases with $\sigma_0$. Quite interestingly, the peak of the ratio grows with the tidal charge $b/M^2$. Moreover, we also show the ratio as a function of the tidal charge $b/M^2$ for $\sigma_0$=100 in Fig. \ref{Ratiobmb}. The X-point location is set to $r/M$=1.3, 1.4, 1.5, 1.6, 1.7 from right to left. For each fixed $r/M$, we observe that the ratio of the power increases with the tidal charge $b/M^2$. Near the maximal value of $b/M^2$, the ratio will be larger than one. Therefore, in some certain parameter values, the power of energy extraction via fast collisionless magnetic reconnection can exceed
that from the BZ process.

\begin{figure}
\center{\includegraphics[width=7cm]{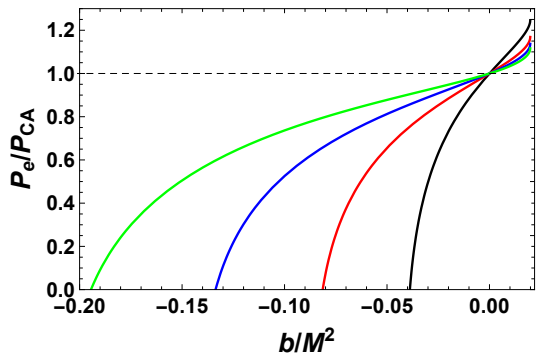}}
\caption{Power ratio $P_{\text{e}}/P_{\text{Kerr}}$ as a function of $b/M^2$ for a rapidly spinning black hole with $a/M$=0.99, $\sigma_{0}=100$, $\xi=\frac{\pi}{12}$, and $r/M$=1.3, 1.4, 1.5, 1.6 from right to left in the bottom.} \label{PRPeKerr}
\end{figure}

In the above, we have considered the ratio between the power of the magnetic reconnection and the BZ mechanism. Here we would like to examine the power ratio $P_{\text{e}}/P_{\text{CA}}$ of the Comisso-Asenjo mechanism between the spinning braneworld black hole and Kerr black hole. The ratio is plotted in Fig. \ref{PRPeKerr} as a function of the tidal charge $b/M^2$. For different values of $r/M$, we find that the ratio increases with $b/M^2$. Moreover, we also observe that the power ratio is smaller than one for negative tidal charge, while large than one for positive tidal charge. This suggests that comparing with the Kerr black hole, energy extraction from a braneworld black hole with positive tidal charge is more efficient via the magnetic reconnection mechanism. In addition, the ratio approaches its maximal values for the extremal black holes. This result is also consistent with the efficiency $\eta$ of the reconnection process shown in Fig. \ref{Effecbb}.

\section{Conclusions and discussions}
\label{Conclusion}

In this paper, we followed the work of Comisso and Asenjo \cite{Comisso} to extract black hole rotational energy via the fast magnetic reconnection for a spinning braneworld black hole, and analytically studied the effects of the tidal charge on this mechanism.

In order to make this process of energy extraction more clear, we would like to give a brief review here. In this mechanism \cite{Comisso}, there are two key points. One is the generation of the magnetic reconnection. The second is that the infalling particle has negative energy-at-infinity and the escaping particle has positive energy. Considering the conservation of energy, extra energy will be carried out by the escaping particle from the black hole. As already known, the phenomenon of the magnetic reconnection can occur near the equatorial plane of the spinning black hole when the magnetic field direction changes caused by the plasmoid instability. In addition, negative energy-at-infinity can exist in the ergosphere of a rotating black hole. These indicate that such energy extraction mechanism is realizable. In particular, via calculating the power and efficiency of this mechanism, the result suggests that it can be comparable with the BZ process  \cite{Comisso}.

Since the magnetic reconnection is required to occur and accelerate the plasma in the black hole ergosphere, we in this paper studied the possible circular orbits near the spinning braneworld black hole in the equatorial plane. Employing the effective potential of the radial motion, we obtained the radii of the outer black hole horizon, ergosphere, light ring and ISCO. For a fixed black hole spin, all of them decrease with the tidal charge $b$. When the maximal value of the tidal charge is reached, all the radii except that of the ergosphere meet at $r/M$=1.

Then we calculated the energy-at-infinity for the accelerated/decelerated plasma, which is dependent of five parameters ($a$, $b$, $r$, $\sigma_0$, $\xi$). For the rapidly spinning black hole, we observed that both energies $\epsilon_{\pm}^{\infty}$ decrease with the tidal charge until $b_{\text{max}}$ is reached. In particular, the starting points exactly locate at the radii of the corresponding light rings. Also, $\epsilon_{+}^{\infty}$ increases and $\epsilon_{-}^{\infty}$ decreases with the plasma magnetization $\sigma_0$. In order to extract the black hole energy, one requires the accelerated plasma has positive energy and the decelerated one has negative energy, i.e., the conditions (\ref{condds}) should be satisfied. Aiming at this, we examined the possible parameter region for a rapidly spinning black hole, which is expected to be a good candidate for energy extraction. As shown in Fig. \ref{ppParaRgepi6}, larger $\sigma_0$ enlarges the region satisfying the conditions. And the region extends to large X-point location $r/M$ and small tidal charge $b/M^2$. However the region shrinks with the orientation angle $\xi$.

The power and efficiency of the energy extraction via the magnetic reconnection for the rapidly spinning braneworld black hole were also calculated. With the increase of the tidal charge $b/M^2$, the power first increases, and then decreases for large X-point location $r/M$. However for small $r/M$, the power monotonically increases until the maximal value of the tidal charge is reached. Meanwhile, the power will be significantly enhanced by the plasma magnetization $\sigma_0$ as expected. The efficiency was also found to increase with the tidal charge, and the maximal efficiency always occurs at the maximal tidal charge. Although large $r/M$ is related with small efficiency, it enlarges the region $\eta>1$ for the tidal charge.

We also compared the power of the magnetic reconnection process and the BZ process. The ratio was numerically calculated. The result suggests that the maximal ratio always happens at low $\sigma_0$. A detailed calculation also shows that the ratio increases with the tidal charge and the maximal ratio is at $b_{\text{max}}$. In particular, in certain parameter regions of the tidal charge, the ratio is higher than one indicating that the power of the magnetic reconnection process is more efficient than the BZ process. So the magnetic reconnection can serve as a visible energy extraction mechanism.

In summary, energy extraction via magnetic reconnection process in the rapidly spinning braneworld black hole was promised. Comparing with the spinning Kerr black hole counterpart, the tidal charge grows the maximal power and efficiency of the energy extraction. Meanwhile the present of the tidal charge can make the magnetic reconnection more efficient than the BZ mechanism.

\section*{Acknowledgements}
This work was supported by the National Natural Science Foundation of China (Grants No. 12075103, No. 11675064, No. 11875151, and No. 12047501), the China Postdoctoral Science Foundation (Grant No.2021M701531), the Fundamental Research Funds for the Central Universities (Grants No. lzujbky-2021-it34, No. lzujbky-2021-pd08), the 111 Project (Grant No. B20063), and Lanzhou City's scientific research funding subsidy to Lanzhou University.

\end{document}